# High-Throughput Density Functional Theory Screening of Double Transition Metal MXene Precursors

## Authors
Kat Nykiel, Alejandro Strachan
**Affiliations**
School of Materials Engineering and Birck Nanotechnology Center, Purdue University, West Lafayette, Indiana 47907, United States

## Abstract
MXenes are an emerging class of 2D materials of interest in applications ranging from energy storage to electromagnetic shielding. MXenes are synthesized by selective etching of layered bulk MAX phases into sheets of 2D MXenes. Their chemical tunability has been significantly expanded with the successful synthesis of double transition metal MXenes. While knowledge of the structure and energetics of double transition metal MAX phases is critical to designing and optimizing new MXenes, only a small subset of these materials been explored. We present a comprehensive dataset of key properties of MAX phases obtained using density functional theory within the generalized gradient approximation exchange-correlation functionals. Energetics and structure of 8,712 MAX phases have been calculated and stored in a queryable, open database hosted at nanoHUB.

## Background & Summary

The recent emergence of atomically-thin materials within the last two decades has precipitated a multitude of new families of 2D materials, each offering complementary properties, including remarkable mechanical, optical, and electronic properties.[1,2] One class of 2D materials with a recent rise in interest is MXenes, a family of layered transition metal carbides and nitrides first discovered in 2011.[3]

The synthesis of MXenes is accomplished by preferential etching of layered MAX phases. These MAX phases consist of *M*, *A*, and *X*-type atoms, where *M* denotes an early transition metal, *A* denotes a group 13-16 element, and *X* denotes carbon or nitrogen. The M, A, and X atoms crystallize into a layered structure, which forms the basis for 2D MXene sheets. Fig. 1 shows the three crystal structures observed in MAX phases. From the MAX phase precursor, the comparatively weaker M-A bonds can be etched away, resulting in single- or multi-layered MXenes. This unique top-down synthesis approach gives MXenes more scalability than other families of 2D materials. Beyond their synthesis route, MXenes are desirable for applications in energy storage, catalysis, and electronic sensors due to their metallic conductivity and 2D structure.[4] For these reasons, MXenes have a large potential in both existing and future applications.

A recently emerging sub-class of MXenes known as double transition metal (DTM) MXenes contain two different transitional metal elements.[5] These DTM MXenes retain the layered structure of mono-M MXenes, with *M'* and *M''* atoms occupying the *M* sites. These MAX phases are further subdivided into in-plane ordered (i-MAX), out-of-plane ordered (o-MAX), and solid-solution (ss-MAX) phases, depending on the distribution of the M' and M'' atoms.[6]

Out of the 8712 possible o-DTM-MAX materials, only a few have been etched into MXenes. Thus, to guide future experimental efforts, we introduce a comprehensive dataset of density functional theory (DFT) results for all possible o-MAX phases, including energetics and structural information.

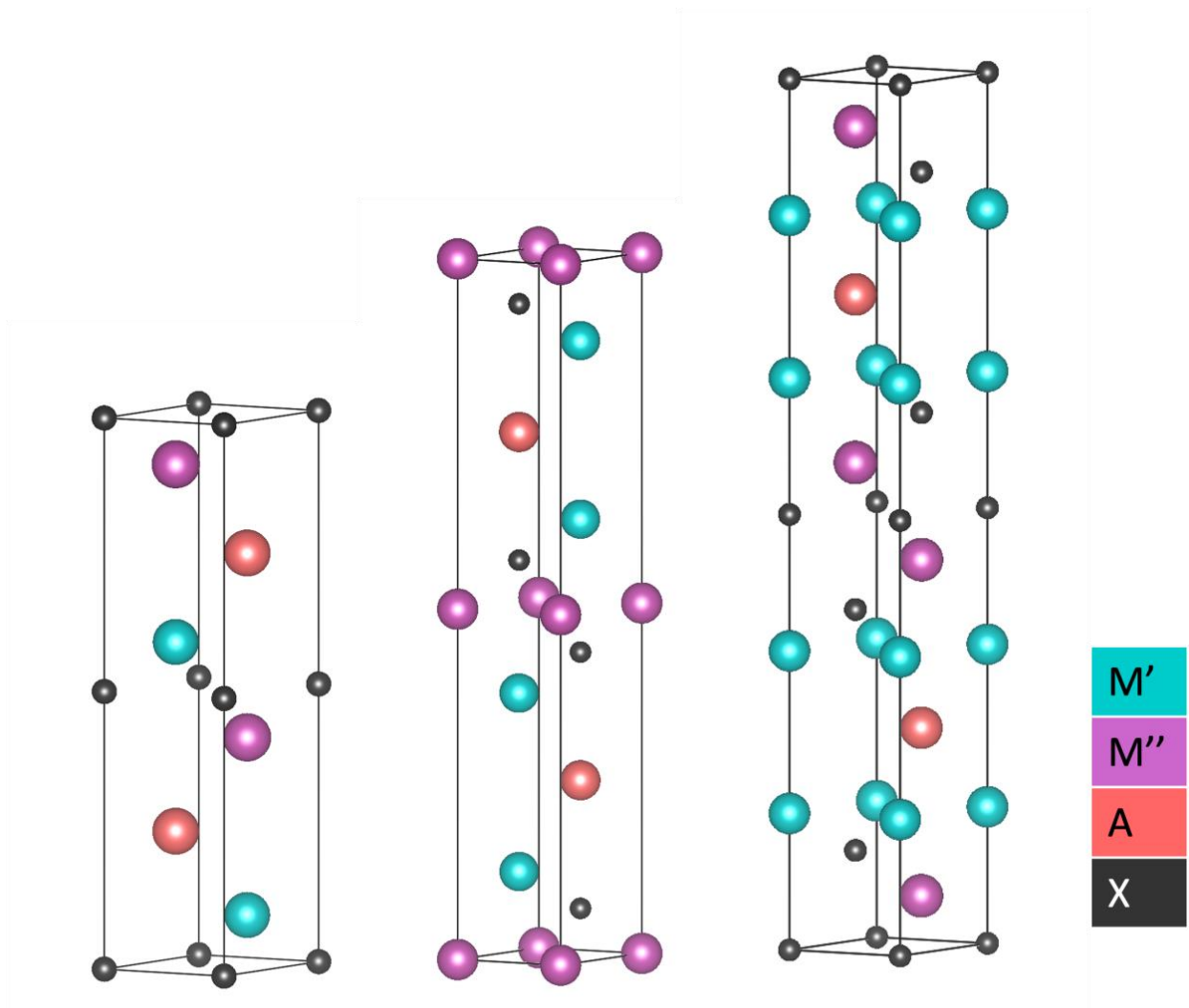

**FIGURE 1: CONVENTIONAL DTM MAX UNIT CELLS, FOR N=1, 2, 3 LAYERED STRUCTURES**

The multivariate composition of DTM MXenes and their precursor MAX phases offer a high-dimensional domain of exploration well-suited for high-throughput density functional theory. Existing databases of DFT calculations, such as Materials Project[7], only contain a small fraction of all possible MAX phases. For example, querying Materials Project's API for all MAX-formulae with space group $P_{63}/mmc$ reveals only 75 MAX phases, none of which are DTM MAX structures.

Several other studies have reported DFT investigations of DTM MAX structures, but none present a complete view of the entire DTM MAX domain of compositions. Exhaustive computational investigations of mono-M MAX phases and their MXenes have been performed[8], but this has not been extended to DTM MAX phases. Several studies have characterized DTM MXenes using DFT, but each was limited in domain to <100 compositions.[9–11] A database of 23,870 DTM MXene phases has previously been established; however, this database contains no precursor MAX phases, and 70% of the MXene properties in the database are machine learning generated.[12]

This paper aims to fill this gap by establishing a complete database of bulk DTM o-MAX structures, which can accelerate the development of new DTM MXenes. Electronic structure calculations using DFT predict several of the properties of interest in MXene applications and help guide the design of new 2D materials of this family. The composition domain of this database is defined below in Fig. 2, where M', M'' = Hf, Mn, Cr, Mo, Nb, Sc, Ti, V, Zr, and Ta, and W; A = Cd, Pb, P, S, Tl, As, Al, Si, Ge, Ga, In, and S; and X=C or N, with n=1, 2, and 3 layers. This database includes 8712 potential MAX phases.

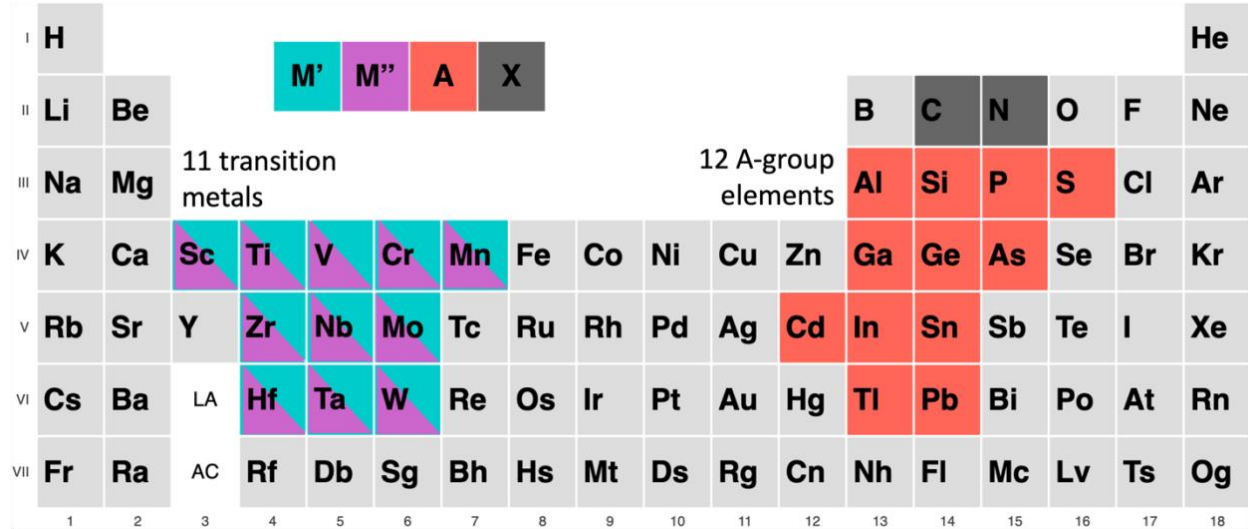

**FIGURE 2: DTM MAX DOMAIN**

## Methods
### Workflow Overview

The data and simulations workflow used to characterize and obtain features for all 8712 candidate MAX phases consists of three main steps, see Fig. 3: i) initial structure generation, ii) DFT calculations, and postprocessing/indexing of results into a queryable database. The following subsections describe each step in detail.

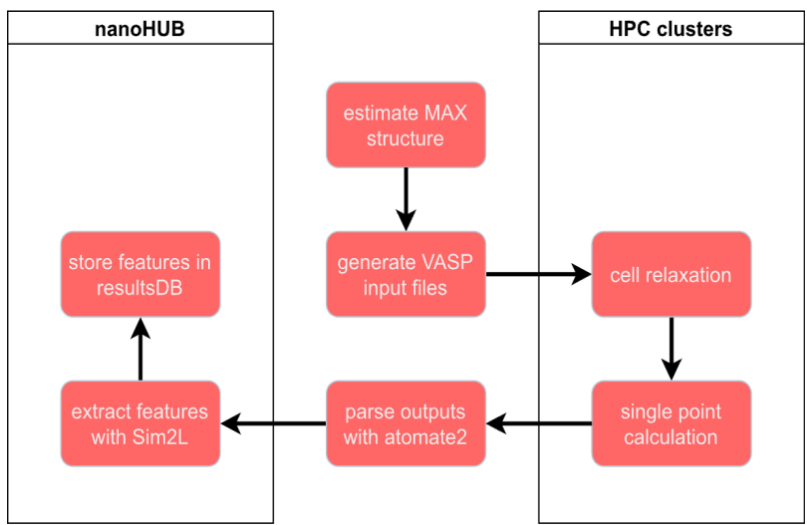

**FIGURE 3: HIGH-THROUGHPUT DFT WORKFLOW**

## Initial Structure Generation

The unit cell for each MAX structure is first estimated starting with the $Ti_3AlC_2$ MAX structure obtained from the Materials Project.[7] From each MAX composition, the selected elements are mapped using pymatgen, and the lattice vectors are adjusted based on the atomic radii of the elements involved, using geometric insight on MAX phases to serve as a starting estimate for the DFT calculations.[13] The expressions used to calculate the a and c lattice parameters for each of the three crystal structures are listed in Tbl. 1.

**TABLE 1: GEOMETRIC ESTIMATES FOR LATTICE PARAMETERS**

| n | a (Å) | c (Å) |
|---|---|---|
| 1 | $2r_{M''}$ | $\frac{4}{\sqrt{2}}(r_{M'} + r_X + 2r_A)$ |
| 2 | $2r_{M''}$ | $\frac{4}{\sqrt{2}}(2r_{M''} + r_{M'} + 2r_X + r_A)$ |
| 3 | $2r_{M''}$ | $4(r_{M'} + 2r_{M''} + r_X + r_A)$ |

Density functional theory calculations were performed on each structure using the Vienna Ab initio Simulation Package (VASP).[14,15] These simulations utilized Perdew–Burke–Ernzerhof (PBE) exchange-correlation functional and projector-augmented wave (PAW) pseudopotentials.[17] The calculations were all spin-polarized. Starting with the initial structures described above, a fixed-cell ionic relaxation via energy minimization is performed. This is followed by a cell relaxation to relax the stress tensor and a single-point energy calculation on this optimized structure to obtain energetic properties. Ionic relaxations use the conjugate gradient algorithm with $10^{-5}$ eV stopping criteria and the cell relaxation uses conjugate gradient algorithm and $10^{-6}$ eV threshold. To ensure numerical convergence of the runs, a kinetic energy cutoff of 550 eV and a k-point grid of 18x18x3 was chosen. This was determined to be a sufficient cutoff by ensuring a variability of less than 1% of the total *a* and *c* lattice parameter values.

## Simulation Results Postprocessing and Indexing into a Queryable Database

The raw VASP outputs are processed to extract the structural and energetics of interest using the following procedure. The first step is to parse the VASP output using the *VaspDrone* function of *atomate2*[18] to convert it into a structured JSON document. Atomate2 is an open-source set of computational materials science workflows designed for high-throughput automation. To extract the desired quantities of interest and make the data Findable, Accessible, Interoperable, and Reusable (FAIR)[19], we use nanoHUB's[20] Sim2Ls.[21] Sim2Ls are online, end-to-end, and queryable workflows with formally declared inputs and outputs. They are implemented as Jupyter notebooks and can be executed from nanoHUB or using an API. The open source *VASPINGESTOR* Sim2L ingests JSON files produced by atomate2 and extracts the desired structural and energetic information.[22] Importantly, all the inputs and outputs, including the JSON file and auxiliary quantities of interest, are automatically indexed into nanoHUB's results database (resultsDB) where they can be accessed through a web interface or queried using an API. The Section **Data Records** describes the properties extracted from the JSON file with the raw results and stored in the ResultsDB. We note that the *VASPINGESTOR* tool is open and available online, it can be used by any researcher using VASP to make their data FAIR.

## Data Records

As described above, we use nanoHUB's Sim2Ls to make both the postprocessing workflow and the data FAIR.[21] All the inputs and the outputs of the vaspingestor Sim2L are indexed in the ResultsDB, they are listed in Tbl. 2 and Tbl. 3 with an (I) or (O) following their names, respectively. The features extracted from the 8712 DFT simulations can be grouped into two categories, the first being those directly extracted from the VASP output, such as energies, stresses, and structures. These results are listed in Table 2.

**TABLE 2: FEATURES STORED IN SIM2L**

| Name | Type | Unit | Description |
|---|---|---|---|
| Doc (I) | Dictionary | n/a | Atomate2 TaskDocument to be converted to a set of interpretable features |
| Author (I) | String | n/a | Name of author associated with this run |
| Dataset (I) | String | n/a | Identifying tag for the dataset being stored |
| Structure (O) | Dictionary | n/a | pymatgen Structure object, containing lattice vectors and atomic positions |
| Composition (O) | Dictionary | n/a | Chemical composition of the unit cell |
| lattice_parameters (O) | Array | Å | a, b, c lattice parameters of unit cell |
| Energy (O) | Number | eV | Total energy of the system |
| Stress (O) | Array | kbar | External pressure of the system |
| Forces (O) | Array | eV/Å | List of (x,y,z) forces on each atom |
| max_force (O) | Number | eV/Å | Maximum force reported during the simulation |
| rms_force (O) | Number | eV/Å | Root mean square force reported during the simulation |
| KPOINTS (O) | Array | n/a | Number of k-points in the x, y, and z directions |
| ENCUT (O) | Number | eV | Kinetic energy cutoff for the plane wave basis set |
| XC_functional (O) | String | n/a | Choice of exchange-correlation functional used, read from VASP's GGA tag |
| Pseudopotential (O) | String | n/a | Choice of pseudopotential used |

From the DFT results, additional quantities of interest useful as descriptors for machine learning applications are extracted, these are provided in Table 3. These quantities are categorized separately as they either require additional DFT calculations, such as formation and cohesive energy, or they are MAX-specific, such as the selected bond lengths and interlayer distances.

**TABLE 3: FEATURES DERIVED FROM VASP RESULTS**

| Name | Type | Unit | Description |
|---|---|---|---|
| Formation energy (O) | Number | eV | Formation energy of the system, calculated using Eq. 1 |
| Cohesive energy (O) | Number | eV | Cohesive energy of the system, calculated using Eq. 2 |
| Bond lengths (O) | Array | Å | Distance between MX and MA atoms within the MAX phase, as defined in Fig. 4 |
| Interlayer distances (O) | Array | Å | Distance between MM, MX, XA, and AA layers within the MAX phase, as defined in Fig. 4 |

The bond lengths and interlayer distances listed in Table 3 are illustrated in Fig. 4.

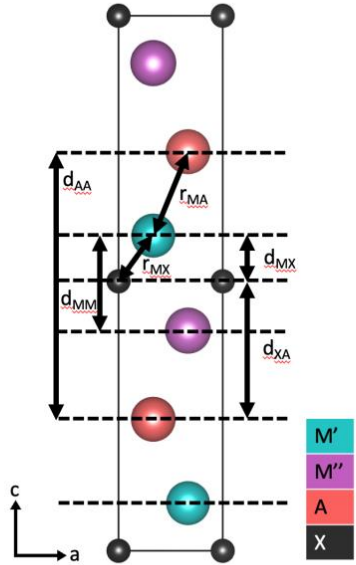

**FIGURE 4: INTERLAYER DISTANCES AND BOND LENGTHS**

Formation energies are calculated using:

$$E_{\text{form}}^{MAX} = E_{\text{bulk}}^{MAX} - \sum_{i=\{M,A,X\}} N_i * \left(e_i^0\right) \quad (1)$$

This equation takes the difference of the obtained MAX phase energy $E_{bulk}^{MAX}$ and the constituent elements $i$ in their respective standard states, $e_i^0$, which represent the energy per atom in the standard state and $N_i$ the number of atoms of type $i$ in the MAX structure. These equilibrium states are the lowest energy configurations of each element; for example, the equilibrium state of carbon would be the graphite structure. This provides a metric of the relative stability of the MAX phase over the stability of the individual components.

Finally, cohesive energies were calculated using Eq. 2.

$$E_{\text{form}}^{MAX} = E_{\text{bulk}}^{MAX} - \sum_{i=\{M,A,X\}} N_i * E_i^{\text{vacuum}} \quad (2)$$

This equation takes the difference of the obtained MAX phase energy $E_{bulk}^{MAX}$ and the constituent elements $i$ as single atoms in vacuum. This provides an additional metric of stability as the difference between a bonded and dissociated structure.

## Data visualization

As an illustration of the depth and range of this dataset, Fig. 5 shows the formation energy of all the structures relaxed as a function of the c lattice parameter. The data points are colored by metallic atom M' (a), A atom (b), carbide or nitride (c), and the number of layers in the structure (d). While the plots colored by X and n demonstrate clear clustering, the plots of M' and A demonstrate trends within each of these colors, such as the shift of lattice parameter in the A-colored plot from smaller radii (S, P) to larger radii (Pb, Cd). The plot colored by X demonstrates the difference between carbide and nitride MAX phases in terms of formation energy. This is further highlighted in Fig. 6 where we use violin plots to formation of carbides and nitrides. The results demonstrate the broad range of formation energies and nitrides being, on average, more stable than carbides. This is interesting given the relative lack of experimentally synthesized N-based MAX phases identified by Hong et al.[6] We, therefore, identify this as a potential avenue for further exploration.

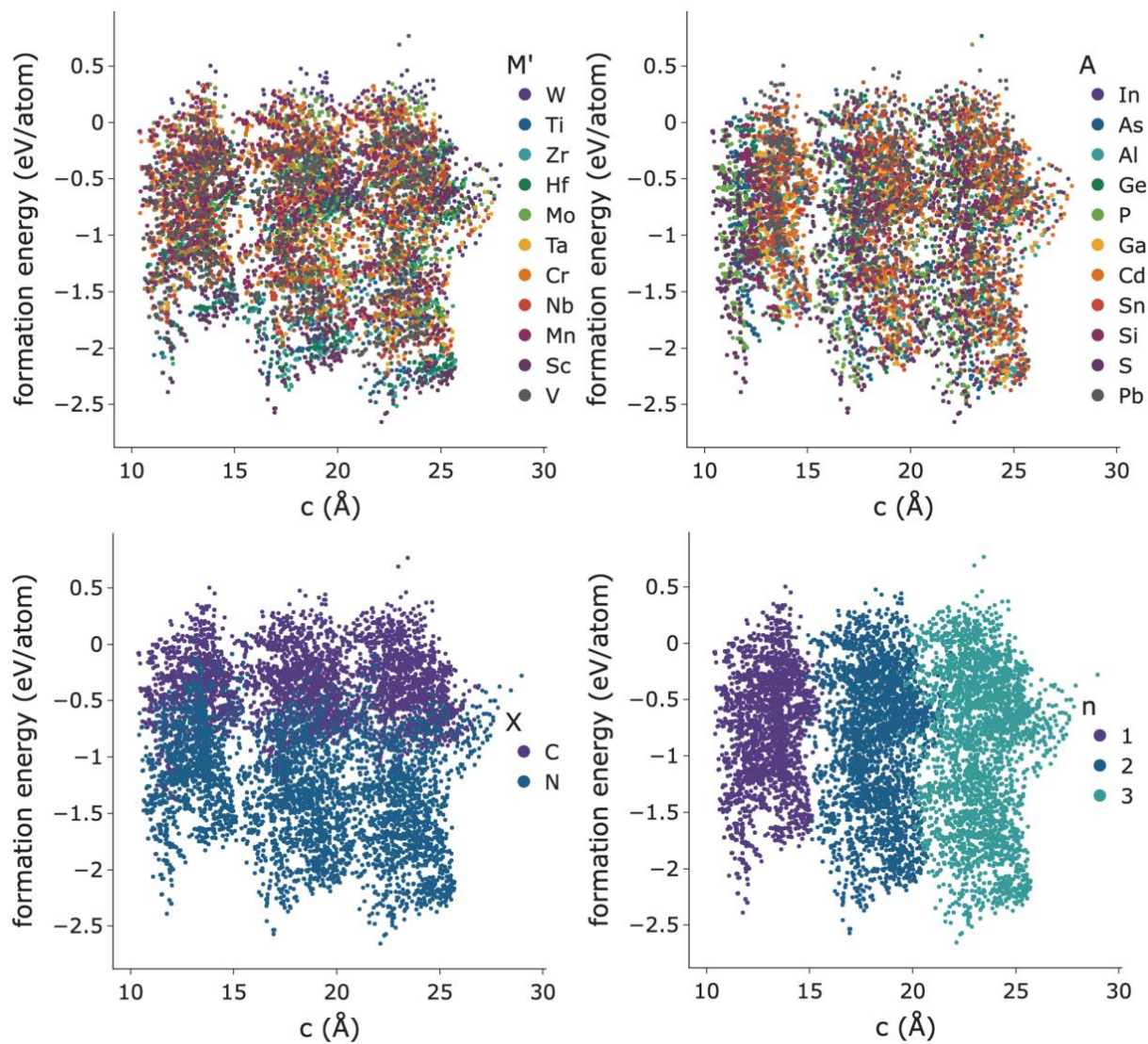

**FIGURE 5: FORMATION ENERGY VS. LATTICE PARAMETER, COLORED BY M', A, X ELEMENTS AND N, NUMBER OF LAYERS**

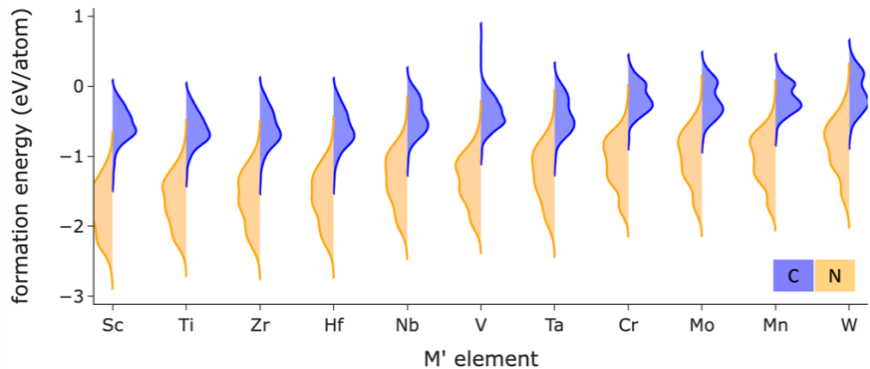

**FIGURE 6: VIOLIN PLOT OF FORMATION ENERGIES FOR C AND N-BASED MAX PHASES AS A FUNCTION OF M'**

The formation energy as a function of M' and M'' metallic elements is explored in Fig. 7 for all the A elements explored. This plot demonstrates the relative formation energy magnitudes of each M'/M'' combination for C-based MAX phases. Generally, N-based MAX phases exhibit the same trends in formation energy as a function of M'/M''. Both C-based and N-based figures are available in the vaspingestor tool on nanoHUB. These slices demonstrate that MAX phases with Group 4 elements (Ti, Zr, Hf) yield the lowest formation energies.

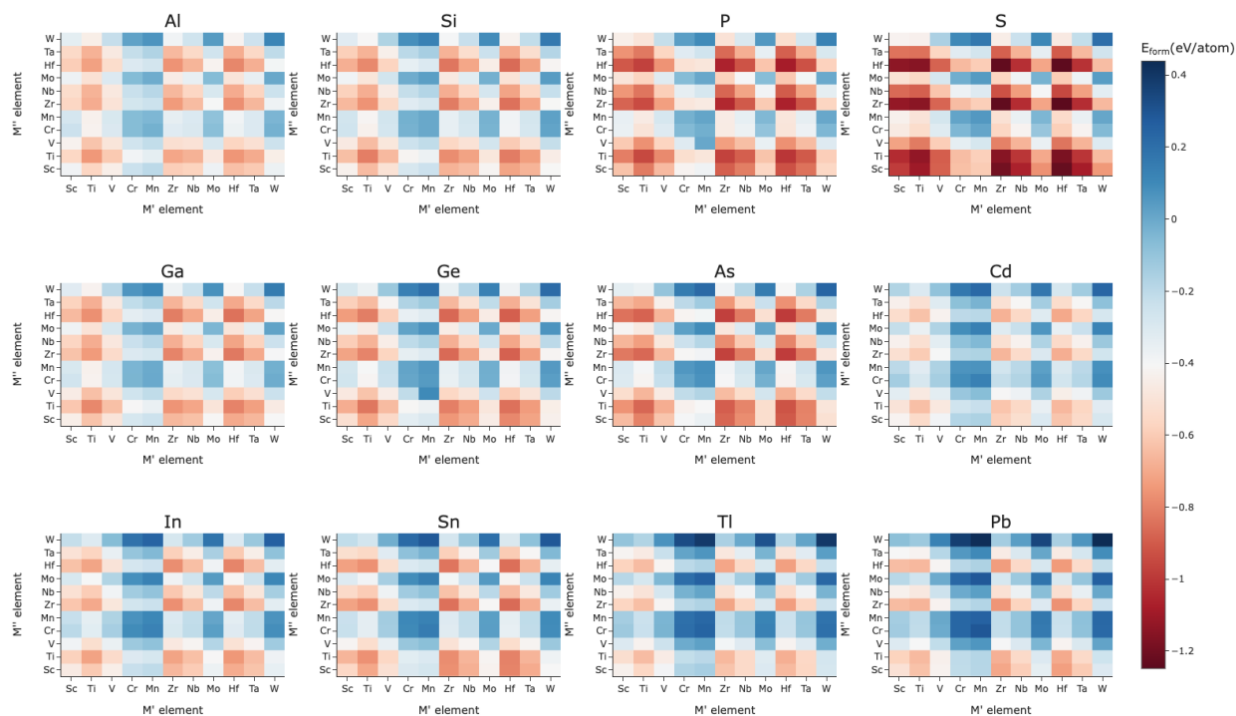

**FIGURE 7: SLICES OF C-BASED MAX DATASET WITH CONSTANT A, TABULATED WITH M'/M'', COLORED BY FORMATION ENERGY.**

Figures 5, 6, and 7 can be re-created by querying the nanoHUB ResultsDB, this is demonstrated in the VASPINGESTOR online tool.[22] The tool also includes an example of how researchers can upload their own VASP results.

## Technical Validation
### Comparison to Frey et al.

The largest existing database of MAX phases, published by Frey et al.[8], contains 792 MAX phases with single-M structures. This work reported lattice parameters, formation energies, and cohesive energies for all calculated MAX phases. A comparison between their MAX phase results and this work is provided in Fig. 8 as parity plots of formation and cohesive energies as well as $a$ lattice parameters. Values are generally in agreement for the formation and cohesive energies, with RMSE values of 0.13 and 0.25 eV/atom, respectively. As shown in Fig. 8, the major outliers are limited to Sc-A-C type MAX phases, where A is any of the 12 possible A-group elements. We independently confirmed our calculations and we compared them with the Sc-containing MAX phases in the Materials Project. Both these tests indicate our results to be correct. A possible origin of this discrepancy is differences in the spin state of the systems. Comparisons of lattice parameters indicate an RMSE of 0.12 Å with 76% of the structures showing discrepancies less than 2%. 10% of the structures show differences of over 5% which could not be explained by differences in the methods used. This comparison and analysis highlight the importance of making research data FAIR, as Frey et al. did.

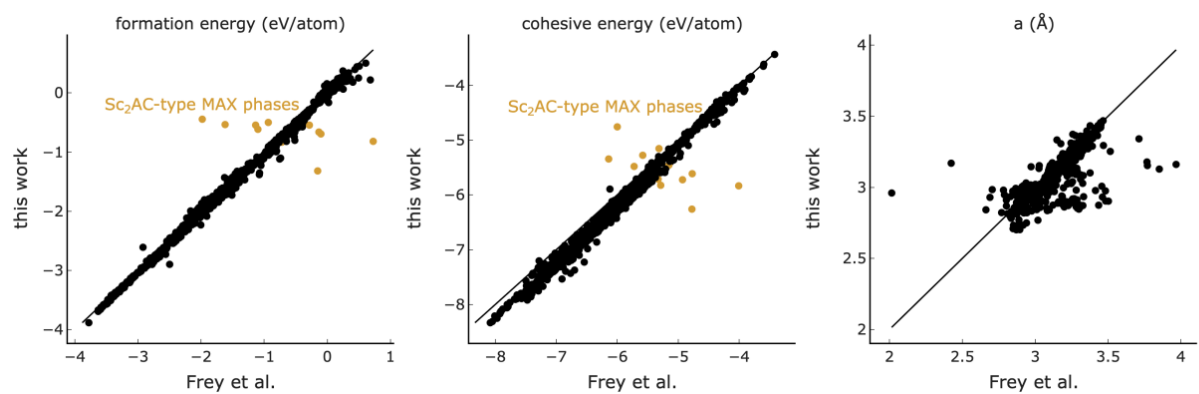

**FIGURE 8: COMPARISON OF LATTICE PARAMETER, FORMATION AND COHESIVE ENERGIES WITH FREY ET AL.**

### Comparison to Materials Project

To further validate our data, a comparison was made to all existing MAX phases in the Materials Project.[7] The formation energies and $a$ lattice parameters are shown in Fig. 9. In the plot of formation energy, we provided reference lines to account for the disparity in nitrogen reference between datasets. We use nitrogen in its gaseous state as its equilibrium structure, while the Materials Project uses a crystalline form of nitrogen as its reference. When accounting for this difference the formation energies and lattice parameter RMSE values were 0.06 eV/atom and 0.07 Å. The Sc-A-C type MAX phase formation energies have an RMSE of 0.004 eV/atom, showing strong agreement with Materials Project and exhibiting the same deviation from Frey et al. as shown in Fig. 8.

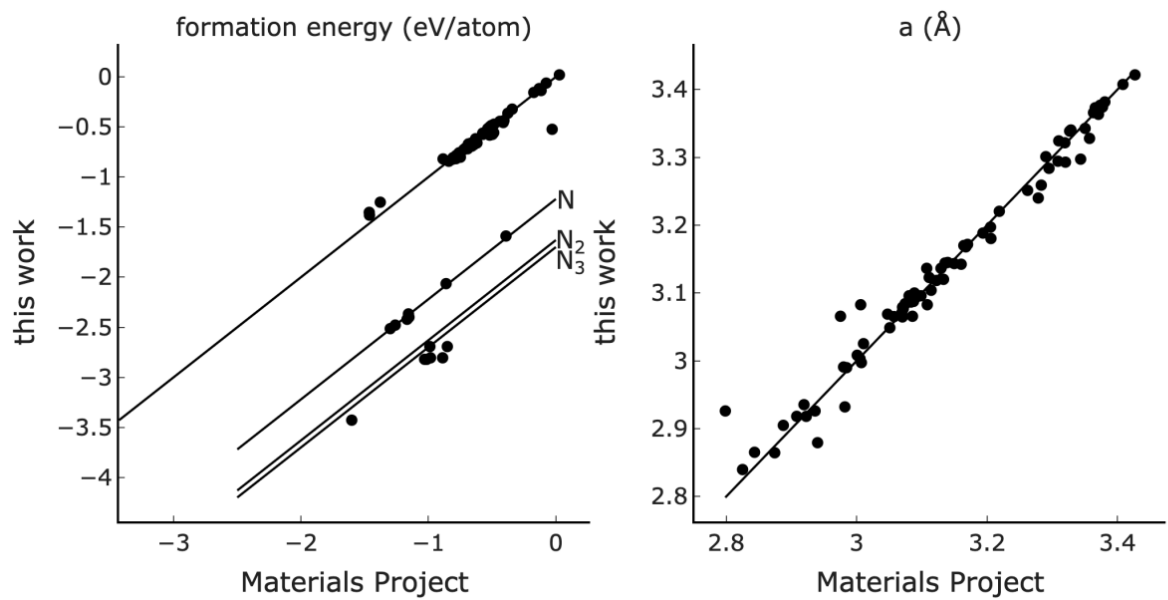

**FIGURE 9: COMPARISON OF FORMATION ENERGIES AND LATTICE PARAMETERS WITH MATERIALS PROJECT**

## Code Availability
The DFT results database, code used to generate figures, and tool for ingesting new data into this database are available at https://nanohub.org/tools/vaspingestor. This code is also available at https://github.com/katnykiel/vasp_ingestor.

## Acknowledgements
We acknowledge support from National Science Foundation, Award 2124241, and nanoHUB.org for storing the data and hosting the tool to access data.

## Author contributions
Kat Nykiel ran the DFT calculations, build the analysis workflow and developed the nanoHUB tool.
Alejandro Strachan guided the workflow, tool, and analysis.

## Competing interests
The authors declare no competing financial interest.

# Figures
Embedded within Word document.

# Figure Legends

1. Conventional DTM MAX unit cells, for n=1, 2, 3 layered structures. These unit cells demonstrate the three crystal structures studied in this investigation and provide the basis for DFT calculations.
2. DTM MAX domain. This graphic shows the composition space of the elements being studied in the MAX system, with separate colors for M', M'', A and X.
3. High-throughput DFT workflow. This flowchart shows the process by which data is passed from an initial starting structure to the final ResultsDB, hosted on nanoHUB.
4. Interlayer distances and bond length. This figure illustrates the bond lengths and interlayer distances provided in the feature set, to avoid ambiguity in their definition.
5. Formation energy vs. lattice parameter, colored by M', A, X elements and n, number of layers. This scatter plot is included to demonstrate the wide domain of MAX phases being studied, and to show trends in data as a function of the primary M, A, and X elements, as well as the crystal structure n.
6. Violin plot of formation energies for C and N-based MAX phases as a function of M'. This violin plot shows the formation energy as a function of M to provide insight on the relative stabilities of MAX phases.
7. Slices of C-based MAX dataset with constant A, tabulated with M'/M'', colored by formation energy. This plot shows individual formation energies of all MAX phases being studied, with a 11x11 grid and 12 subplots, each showing the formation energy.
8. Comparison of lattice parameter, formation and cohesive energies with Frey et al. This plot shows the deviations in formation and cohesive energies are mainly limited to Sc-A-C-type MAX phases with n=1, while the lattice parameters show much more variation.
9. Comparison of formation energies and lattice parameters with Materials Project. This plot shows that our calculated lattice parameters agree with Materials Project, and that correcting for the different nitrogen reference state demonstrates agreement in formation energy as well.

# Tables
Embedded within Word document.